\documentclass[twocolumn,showpacs,aps,prl,superscriptaddress,floatfix]{revtex4}
\usepackage{graphicx}
\usepackage{dcolumn}
\usepackage{multirow}
\usepackage{amsmath}
\usepackage{epsfig}
\usepackage{color}
\usepackage{subfigure}

\input pubboard/babarsym

\newcommand{\BABARPubYear}    {06}
\newcommand{\BABARPubNumber} {067}
\newcommand{\SLACPubNumber}{12442}

\def\radbhabha{\ensuremath {e^+e^-\gamma}\xspace}
\def\raddimuon{\ensuremath {\mu^+\mu^-\gamma}\xspace}

\def\btoclnu{\ensuremath {{b} \rightarrow c \ell \nu_{\ell} }\xspace}
\def\btoulnu{\ensuremath {{b} \rightarrow u \ell \nu_{\ell} }\xspace}

\def\bulnu{\ensuremath {{B} \rightarrow X_{u} \ell^{+} \nu_{\ell} }\xspace}

\def\hzlnu{\ensuremath {{ B^{+}} \rightarrow h^{0} \ell^{+} \nu_{\ell} }\xspace}

\def\hplnu{\ensuremath {{ B^{0}} \rightarrow h^{-} \ell^{+} \nu_{\ell} }\xspace}

\def\rhopm{\ensuremath {\rho^{\pm}}\xspace}
\def\rhoz{\ensuremath {\rho^{0}}\xspace}

\def\munugamma{\ensuremath {{ B^{+}} \rightarrow \gamma \mu^{+} \nu_{\mu}}\xspace}

\def\nuep        {\ensuremath{\Delta_{EP}}\xspace}

\def\dzeropifull{\ensuremath { B^{+} \rightarrow \pi^{+}\Dzb (\rightarrow K^{+}\pi^{-})}\xspace}

\def\mb        {\mbox{$m_{\rm ES}$}\xspace}

\def\delbf        {\mbox{$\Delta\mathcal{B}$}\xspace}

\def\coslg{\mbox{$\cos\theta_{\ell\gamma}$}\xspace}
\def\CTBLG{\mbox{$\cos\theta_{BY}$}\xspace}

\def\fisher{\ensuremath{\mathcal{F}}\xspace}

\def\BlifetimeRatio{\ensuremath{\tau_{B^\pm}/\tau_{B^0} = 1.071 \pm 0.009}\xspace}
\def\fpmfzz{\ensuremath{1.020 \pm 0.034}\xspace}

\begin{document}

\preprint{\babar-PUB-\BABARPubYear/\BABARPubNumber} 
\preprint{SLAC-PUB-\SLACPubNumber} 

\begin{flushleft}
\babar-PUB-\BABARPubYear/\BABARPubNumber\\
SLAC-PUB-\SLACPubNumber\\
\end{flushleft}

\begin{flushright}
\end{flushright}
\title{\large\bf Search for the Radiative Leptonic Decay $B^{+} \rightarrow \gamma \ell^{+}\nu_{\ell}$}

%
\author{B.~Aubert}
\author{M.~Bona}
\author{D.~Boutigny}
\author{Y.~Karyotakis}
\author{J.~P.~Lees}
\author{V.~Poireau}
\author{X.~Prudent}
\author{V.~Tisserand}
\author{A.~Zghiche}
\affiliation{Laboratoire de Physique des Particules, IN2P3/CNRS et Universit\'e de Savoie, F-74941 Annecy-Le-Vieux, France }
\author{E.~Grauges}
\affiliation{Universitat de Barcelona, Facultat de Fisica, Departament ECM, E-08028 Barcelona, Spain }
% ...
\collaboration{The \babar\ Collaboration}
\noaffiliation

\date{\today}

\begin{abstract} 
We present the results of a search for $B^{+}\rightarrow\gamma \ell^{+}\nu_{\ell}$, 
where $\ell = e, \mu$. 
We use a sample of 232 million $B\bar{B}$ pairs recorded at the $\Upsilon(4S)$ 
with the \babar\ detector at the PEP-II $B$~Factory.
We measure a partial branching fraction \delbf in a restricted region 
of phase space that reduces the effect of theoretical uncertainties,
requiring the lepton energy to be between $1.875$ and $2.850\gev$,
the photon energy to be between $0.45$ and $2.35\gev$, and
the cosine of the angle between the lepton and photon momenta to
be less than $-0.36$, with all quantities computed in the \Y4S
center-of-mass frame.  We find
 $\Delta\mathcal{B}(B^{+}\rightarrow\gamma \ell^{+}\nu_{\ell}) = 
(-0.3^{+1.3}_{-1.5}(\mbox{stat})\pm0.6(\mbox{syst})\pm0.1(\mbox{th}))\times10^{-6}$, 
assuming lepton universality.
  Interpreted as a 90\% C.L.
Bayesian upper limit, the result corresponds to $1.7\times10^{-6}$ 
for a prior flat in amplitude, and $2.3\times10^{-6}$
for a prior flat in branching fraction.
\end{abstract}


\maketitle  

At tree level, the branching fraction (BF) for radiative leptonic $B$ decays is
given by:
\begin{equation}
\label{eqn:first}
\mathcal{B}(B^{+} \rightarrow \gamma\ell^{+}\nu_{\ell}) = \alpha\frac{G^{2}_{F} |V_{ub}|^{2}}{288\pi^2}f^{2}_{B}\tau_{B}m_{B}^5\left(\frac{Q_u}{\lambda_{B}} - \frac{Q_b}{m_b}\right)^2,
\end{equation}
where $m_{B}$ is the $B^{+}$ meson mass, 
$m_{b}$ is the $\overline{M\!S}$ $b$ quark mass, $\tau_{B}$ is 
the $B^{+}$ meson lifetime, $f_B$ is the $B$ meson 
decay constant, $Q_i$ is the charge of quark flavor $i$ and 
$\lambda_{B}$ is the first inverse moment of the $B$ light-cone
distribution amplitude~\cite{korchemsky,ref:genon}, a quantity
that enters into
 theoretical calculations~\cite{ref:beneke} of the 
BF of hadronic $B$ decays such as $B\rightarrow \pi\pi$, and
is typically taken to be of the order of $\Lambda_{\mathrm{QCD}}$.
Thus, a measurement of $\mathcal{B}(B^{+} \rightarrow \gamma\ell^{+}\nu_{\ell})$
can provide a determination of
$\lambda_{B}$ free of hadronic final-state uncertainties.
The best current 90\% C.L. upper limit on the full BF is $5.2\times10^{-5}$\cite{ref:cleo}, for
\munugamma.

However, Eq.(1) is based on the assumption that the factorization relation 
for the vector and axial-vector form factors is valid over the entire phase space.  
Instead, one can relate, at tree-level, $\lambda_B$ to a partial BF, \delbf, over
a restricted region of phase space~\cite{pirjol}:
\begin{equation}
\label{eqn:pirjol}
\delbf = \alpha \frac{G_{F}^2|V_{ub}|^2}{32\pi^4}f_{B}^2\tau_{B} m_{B}^3\left[a + bL + cL^2\right],
\end{equation} 
where $L= (m_B/3)(1/\lambda_B + 1/(2m_b))$, 
the first term describes the effects of photon radiation from the lepton, the third term
the internal photon emission, and second their interference.
The constants $a$, $b$, and $c$ can be predicted model-independently using
factorization at large photon energy, the kinematic region for our analysis.

We present herein 
the results of a search for charged $B$ meson decays $B^{+} \rightarrow \gamma\ell^{+}\nu_{\ell}$, where $\ell = e , \mu$ (``electron channel'',``muon channel'')\cite{ccmodes}.
Our measurements are based on a sample of 
232 million \BB\ pairs recorded with the \babar\ detector~\cite{ref:babar} 
at the PEP-II asymmetric-energy \epem storage rings, comprising an integrated luminosity of 210.5~\invfb
collected at the \FourS resonance (``on-peak''). We also use 21.6~\invfb 
recorded approximately 40 \mev\ below 
the \FourS (``off-peak'').

The analysis procedure consists of selecting a lepton and photon recoiling
against a reconstructed $B$, and identifying signal candidates by
reconstructing the neutrino using missing energy and momentum.
We use a variety of selection criteria, optimized using Monte Carlo (MC)
samples, to discriminate signal from background.
We then extract the number of signal events in data using a binned maximum-likelihood (ML) fit.

The backgrounds are divided into three categories: continuum (non-\BB), specific exclusive \btoulnu 
decays, 
and ``generic $B$'' decays,
defined as a combination of all $B$ hadronic decays, \btoclnu decays, and the remaining 
inclusive \btoulnu decays.
In particular, we study the seven exclusive \btoulnu modes:
\hzlnu $(h^{0} = \pi^0,\rho^0,\eta,\eta',\omega)$ 
and \hplnu $(h^{-} = \pi^{-},\rho^{-})$, referred to below, for each $h$, as the 
``$h$ mode''.

Our signal MC samples were generated using the tree-level model of Ref.~\cite{korchemsky}. 
The $\piz$ and $\pipm$ mode samples were generated using the form factor parameterization
of Ref.~\cite{bk}, with the value of the shape 
parameter based on lattice QCD results~\cite{lqcd:fnal}.
Light cone sum rule-based form factor models were used to generate samples for the 
 $\rhoz$, $\rhopm$, and $\omega$ modes~\cite{lcsr:rho}, and 
$\eta$ and $\eta'$ modes~\cite{lcsr:pi}.

We find an excess of events in the off-peak data compared to 
continuum MC ($\epem \rightarrow \qqbar, \tautau$, and in the muon channel, $\raddimuon$),
with the excess more pronounced in the electron channel.  This is likely
to result from unmodeled higher-order QED and hadronic two-photon events.
We thus use off-peak data instead of continuum MC to represent continuum background 
in our analysis.

We take as the signal lepton and photon 
the highest center-of-mass (CM) energy electron (muon) and the 
highest CM energy photon candidate in each event.
The remaining charged tracks, each assigned a pion mass,
and neutral clusters, treated as photons,
are assigned to the ``recoil $B$'' candidate.  
We reconstruct the recoil $B$ in two ways:
we construct an ``unscaled'' recoil momentum as the sum of
the CM 3-momenta of its constituents, and we
define a ``scaled'' recoil momentum in the direction of the
unscaled recoil, with its magnitude determined 
from the CM energy of the \FourS  and the $B^{\pm}$ mass.
Using either the scaled or unscaled momentum, we
reconstruct the 3-momentum of a corresponding scaled or unscaled 
signal neutrino candidate.  The reconstructed neutrino CM energy 
is calculated as the difference between the CM 
beam energy and the sum of the lepton and photon candidates' CM energies.

We optimize a set of selection criteria for the best signal sensitivity 
at a significance of $3\sigma$
using MC 
samples, 
splitting each in half, with 
one sample used for the optimization and the other used to
evaluate its performance.

On the signal side, we require that the electron (muon) have a CM energy between
$2.00$ and $2.85$ ($1.875$ and $2.775$)\gev. 
We require that the photon have a CM energy
between $0.65$ and $2.35$ ($0.45$ and $2.35$) \gev.
We define \coslg to be the cosine of the angle 
between the lepton and photon in the CM frame, and require its value
to be less than $-0.42$ ($-0.36$).  
We require $-1.10 (-1.05) < \CTBLG < 1.10 (1.00)$, where
\CTBLG is the cosine of the angle between the signal
$B$ and the lepton-photon combination $Y$ in the CM frame~\cite{pirho}, computed from the
known $B$ mass, the beam energy, and the 3-momenta of the signal lepton and photon.

In order to reduce background from neutral hadrons, we require the
lateral moment~\cite{LAT} of the electromagnetic calorimeter energy distribution of the 
signal photon candidate to be less than $0.55$ for both channels.
The polar angle of the photon candidate in the laboratory
 frame is required to be between
$0.326$ and $2.443$ rad for both channels.  
We pair the candidate with every other neutral cluster in the event
and reject events with a pair invariant mass in the \piz mass range 
$123$--$147$ ($116$--$148$) \mev.

We require the difference between the total CM energy of the recoil $B$ constituents
 and the CM beam energy to be between $-5.0$ and $0.9$ ($-2.5$ and $0.7$) \gev.
For the neutrino reconstruction, we require that both the scaled and unscaled 
neutrino
 polar angle in the laboratory frame be between $0.300$ and $2.443$ rad for both channels.

To reduce continuum background, we require the ratio of the
second to zeroth Fox-Wolfram moment~\cite{foxwolf} of all
charged tracks and neutral clusters to be less than $0.5$,
and the absolute value of the
cosine of the angle between the CM thrust axes of the recoil $B$ and the lepton-photon system be
less than $0.98$ ($0.86$).
We use a Fisher discriminant, $\fisher \equiv a_{0}L_{0} + a_{2}L_{2}$,
calculated from the momentum-weighted 
zeroth and second Legendre moments, $L_{0}$ and 
$L_{2}$, of the recoil $B$ about the lepton-photon CM thrust axis,
with coefficients $a_{0}$ and $a_{2}$ equal to $0.43$ and $-1.86$ ($0.008$ and $-1.590$), respectively. $\fisher$ is required to be greater than 1.50 (0.310).

In the electron channel, we veto two-photon events
via the charge-angle correlation of the signal lepton arising from the initial state.
For a positively (negatively)-charged signal electron, we require 
the cosine of its CM polar angle to be between
$-0.74$ and $0.78$ ($-0.94$ and $0.70$). In the muon channel, we require this variable
to be between $-1.00$ and $0.78$ for both charges.  
These criteria were optimized on a loosely-selected sample of events,
where the off-peak data are used for the continuum, and the MC for the signal
and other backgrounds.

We also reject two-photon events using a parameterized 
combination of the missing CM momentum in the beam direction and the invariant mass
of the hypothetical two-photon system.
For the muon channel, the entire observed event is taken as the two-photon system,
while for the electron channel, the signal electron is assumed to be from the 
initial state, and so is excluded from the two-photon system.
The selection criterion was adjusted to preserve a 94\% efficiency for signal
for both channels.

\begin{figure}[]
     \vspace{-.4cm}
     \centering
         {\includegraphics[width=0.5\textwidth]{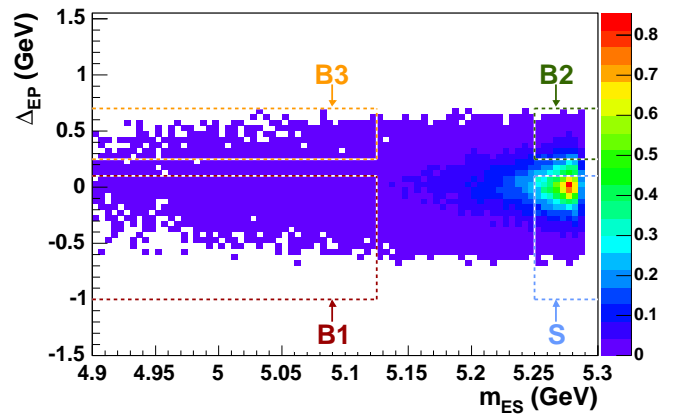}}
     \vspace{-0.5cm}

     \caption{Electron-channel \nuep vs.\ \mb signal MC, using a color scale to 
represent relative contents of each bin.}
     \label{fig:LNG_nuEP_MES}
     \vspace{-.5cm}
\end{figure}

After applying our selection criteria, we use the two-dimensional distribution 
of \nuep,  the difference between the scaled neutrino candidate's CM energy and the 
magnitude of 
its 3-momentum, and \mb, the invariant mass of the recoil $B$, calculated from 
its unscaled CM 3-momentum and 
the CM beam energy, as inputs to the ML fit.  These distributions
provide distinct signatures for signal, $B$ background, and continuum,
with the signal distribution shown in Fig.~\ref{fig:LNG_nuEP_MES}.
The signal (S) and three sideband (B1, B2, B3) regions were selected to maximize
separation of signal from \BB and continuum background.

We extract signal events by fitting on-peak data for the
contributions of signal and background, while allowing the predicted shapes of 
signal and background to vary within statistical uncertainties.
The scale of signal and generic $B$ contributions are allowed to vary,
while the scale of off-peak data is fixed using the on-peak/off-peak luminosity ratio.   
For the seven semileptonic (SL) modes, we
fit for three of the BFs and relate the other four to them as follows:
The $\pipm$ and $\rhopm$ mode BFs are obtained from \babar\ measurements~\cite{pirho}, and
the $\eta$ mode BF is obtained from CLEO~\cite{cleoeta}.
The charged and neutral $\pi$ and $\rho$ modes are
related by the lifetime ratio, \BlifetimeRatio~\cite{pdg2006},
 and an isospin factor of 2.
The $\omega$ mode BF is taken as equal to the $\rhoz$ mode BF.
We take the ratio of the $\eta$ to $\eta'$ mode BFs to be 
$2.057 \pm 0.020$\cite{etaetaprime}.

We maximize a likelihood function consisting of the product of four Poisson
probability distribution functions (PDFs), modeling the 
total counts in each of the four regions, three Gaussian PDFs for
the BFs of the three SL modes, and 40 Poisson PDFs
for the 4-region shapes of the various samples.
All of the shapes are obtained from MC, except for continuum,
where off-peak data are used, introducing a larger statistical uncertainty.

Each Poisson PDF that models the total count in one of the four fit regions has
a measured value obtained from the on-peak data count, and an expected value
based on the fitted contributions of signal and background, including fitted
variations of the shapes.  For the seven SL modes (where three of the fitted BFs are independent), the 
variances of the three Gaussian likelihoods are obtained from
the published statistical and experimental systematic uncertainties, combined in quadrature.
In all, there are 47 PDFs, and 45 free parameters.

We fit for the partial BF \delbf for the kinematic region with
lepton CM energy between $1.875$ and $2.850\gev$, photon CM energy between
$0.45$ and $2.35\gev$, and \coslg 
less than $-0.36$ --- the union of the electron and muon channel regions. 
We perform three fits: separate electron and muon channel fits, and 
a joint fit in which the signal and three SL 
BFs are constrained to be equal for the two channels. For each fit,
errors on the fitted signal BF are obtained by finding the two values
at which 
the signal BF likelihood decreased by a factor of $e^{-1/2}$.
Table~\ref{table:LNG_fit} shows the results from the joint fit.

\begin{table}[h]
\caption{Comparison of fit results and experimental observations for the joint fit to
the muon and electron channels. For each of the four fit regions, the
individual fitted contributions from continuum (cont.), \BB background, and signal are shown,
along with their total. The on-peak and off-peak (scaled to the integrated on-peak
luminosity) observations are shown for comparison, in indented rows, and
are not included in the ``Total fit'' value shown.
}
\vspace{-0.2in}
\label{table:LNG_fit}
\begin{center}
\begin{tabular}{lrrrr} \hline \hline
\multicolumn{5}{c}{Muon channel} \\ \hline
   & S & B1 & B2 & B3 \\ \hline
Fit cont.       &$20.0\!\pm\!11.8$            &$116.3\!\pm\!14.7$          &$42.6\!\pm\!12.8$            &$213.2\!\pm\!42.1$ \\
\quad Off-peak  &$23.0\!\pm\!16.2$            &$158.1\!\pm\!40.8$          &$17.4\!\pm\!12.3$            &$219.7\!\pm\!45.8$ \\
Fit \BB         &$59.1\!\pm\!\phantom{0}8.5$  &$61.0\!\pm\!\phantom{0}9.9$ &$61.7\!\pm\!\phantom{0}9.8$  &$286.6\!\pm\!46.6$\\
Fit signal      &$-5.2\!\pm\!13.8$            &$-1.3\!\pm\!\phantom{0}3.4$ &$-0.4\!\pm\!\phantom{0}1.0$  &$-0.2\!\pm\!\phantom{0}0.5$ \\
\hline
Total fit       &$74.0\!\pm\!\phantom{0}8.1$  &$176.0\!\pm\!12.4$          &$103.9\!\pm\!\phantom{0}9.8$           &$500.0\!\pm\!22.1$ \\
\quad On-peak   &$73.0\!\pm\!\phantom{0}8.5$  &$170.0\!\pm\!13.0$          &$111.0\!\pm\!10.5$           &$498.0\!\pm\!22.3$\\
\hline
& & & &\\
\multicolumn{5}{c}{Electron channel} \\ \hline
   & S & B1 & B2 & B3 \\ \hline
Fit cont.       &$55.4\!\pm\!20.5$            &$181.1\!\pm\!16.2$          &$48.9\!\pm\!14.1$            &$356.7\!\pm\!54.4$ \\
\quad Off-peak  &$41.4\!\pm\!20.7$            &$239.7\!\pm\!48.9$          &$79.0\!\pm\!27.9$            &$294.5\!\pm\!52.9$ \\
Fit \BB         &$69.2\!\pm\!\phantom{0}8.5$  &$59.2\!\pm\!\phantom{0}8.5$ &$140.1\!\pm\!15.5$           &$393.8\!\pm\!57.2$ \\
Fit signal      &$-8.4\!\pm\!22.3$            &$-1.5\!\pm\!\phantom{0}3.9$ &$-1.2\!\pm\!\phantom{0}3.3$  &$-0.4\!\pm\!\phantom{0}1.0$ \\
\hline
Total fit       &$116.2\!\pm\!10.3$           &$238.7\!\pm\!14.5$          &$187.7\!\pm\!12.5$           &$750.2\!\pm\!26.5$ \\
\quad On-peak   &$119.0\!\pm\!10.9$           &$231.0\!\pm\!15.2$          &$176.0\!\pm\!13.3$           &$764.0\!\pm\!27.6$\\
\hline
\hline
\end{tabular}
\end{center}
\end{table}

Table~\ref{table:systematics} shows all systematic uncertainties on \delbf except
for theoretical uncertainties on the signal model, which are shown
in Table~\ref{table:delbf_summary_full}.

The experimental systematic errors result from uncertainties on the data/MC
consistency with respect to tracking efficiency of the signal lepton, 
particle identification efficiency of the signal lepton,
reconstruction of the signal photon energy, 
selection criteria efficiency uncertainties, and uncertainties
on the data/MC consistency in the shape of \nuep and 
\mb.  All of these were evaluated using a number of control samples,
including \raddimuon, \radbhabha, and \dzeropifull.

The uncertainty on the number of produced \BB\ pairs is 1.1\%. 
In our fits, we further assumed a charged-to-neutral $B$ production ratio of 1.0, and
determine the systematic uncertainty by varying within the measured interval
\fpmfzz~\cite{pdg2006}.

For systematic errors due to the theoretical uncertainties on the
$\pi$ and $\rho$ mode BFs and
form factor models, we refit, applying correlated variations in the BFs and form factor models, 
and take the magnitude of the largest change in signal BF as the associated systematic.

For systematic errors arising from the theoretical uncertainties on the $\eta$ and $\eta'$ mode BFs, 
we vary the assumed BFs by $\pm10\%$\cite{cleoeta} and refit
to obtain a systematic error.
We find a negligible systematic from 
the uncertainty on the ratio of the $\eta$ to $\eta'$ mode BFs.

In the generic $B$ sample, a significant fraction of events 
contributing to our fit are 
non-resonant \hbox{\bulnu} events.  We obtain the systematic error due to uncertainty on the \bulnu BF 
by fixing the total contribution of generic $B$ decays, as predicted by MC.

\begin{table}[h]
\caption{Systematic uncertainties on \delbf.
All additive systematic values have been multiplied by $10^{6}$.}
\label{table:systematics}
\begin{center}
\begin{tabular}{lrrr} \hline \hline
Multiplicative   & Muon & Electron & Joint \\ \hline
Tracking efficiency & 1.3\% &1.3\% & 1.3\%\\ 
Particle ID & 3.5\% & 2.2\% & 2.1\%  \\ 
Neutral reconstruction & 1.6\% & 1.6\% &  1.6\% \\ 
Selection efficiency & 6.0\% & 5.0\% &  6.0\% \\  
$B$ counting & 1.1\% & 1.1\% & 1.1\%\\ 
Charged to neutral $B$ ratio & 9.4\% & 9.4\% &  9.4\%\\ 
\hline
Additive   & \multicolumn{3}{c}{} \\ \hline
Shape of \nuep vs.\ \mb &0.3 & 0.2 & 0.3\\ 
$\eta$ mode BF &0.3 &0.1 & 0.2\\ 
$\pi$,$\rho$ mode BF, ff & 0.3 & 0.4 & 0.4\\ 
\bulnu BF  &0.4  &0.2 & 0.3\\ 
\hline
\hline
\end{tabular}
\end{center}
\end{table}

The theoretical uncertainty within the kinematic region of \delbf is 
conservatively estimated by evaluating the change in efficiency
when the model of Ref.~\cite{korchemsky} is modified by setting the axial
vector form factor equal to zero.

The results for \delbf are given in 
Tables~\ref{table:delbf_summary_full} and \ref{table:delbf_ul_summary_full}.
We determine 90\% C.L. Bayesian upper limits by integrating the signal BF likelihood with
 two different priors, both of which take values of 0
for negative values of the signal BF:
a prior flat in the BF (``flat BF prior''), 
and a prior flat in the square root of the BF (``flat amplitude prior''),
equivalent to assuming a flat prior for $|V_{ub}|$ or $f_B$.

\begin{table}[h]
\caption{Comparison of \delbf two-sided results 
for all three fits.  \hbox{All values} have been multiplied by $10^{6}$.}
\label{table:delbf_summary_full}
\begin{center}
\begin{tabular}{lrrrr} \hline \hline
All                    & Central & Statistical & Systematic  & Theoretical\\   
values $\times10^{6}$  & value & uncertainty & uncertainty & uncertainty\\   
\hline
\multirow{2}{*}{Muon} & \multirow{2}{*}{$ -1.33$}  & $+1.74 $ & $+0.80$  & \multirow{2}{*}{$0.03$}   \\ 
                      &                            & $-2.20 $ & $-0.87$  &  \\ 
\multirow{2}{*}{Electron} & \multirow{2}{*}{$ 0.11$}  & $+1.73 $ & $+0.61$  & \multirow{2}{*}{$0.08$}   \\ 
                      &                               & $-2.13 $ & $-0.59$  &  \\ 
\multirow{2}{*}{Joint} & \multirow{2}{*}{$ -0.25$}  & $+1.33 $ & $+0.60$  & \multirow{2}{*}{$0.07$}   \\ 
                      &                               & $-1.53 $ & $-0.64$  &  \\ 
\hline\hline
\end{tabular}
\end{center}
\vspace{-0.3in}
\end{table}
\begin{table}[h]
\caption{The $90\%$ Bayesian upper-limits for all three fits, for the two different choices of prior, 
in terms of \delbf.}
\label{table:delbf_ul_summary_full}
\begin{center}
\begin{tabular}{lrr} \hline \hline
  & Prior flat in amplitude & Prior flat in BF\\ 
\hline
Muon & $<1.5\times10^{-6}$ &  $<2.1\times10^{-6}$ \\ 
Electron & $<2.2\times10^{-6}$ &  $<2.8\times10^{-6}$\\ 
Joint & $<1.7\times10^{-6}$ &  $<2.3\times10^{-6}$\\ 
\hline\hline
\end{tabular}
\end{center}
\vspace{-0.2in}
\end{table}

For our kinematic region, the constants $a$, $b$, and $c$ of
Eq.(2) are $0.88$, $-3.24$, and $3.25$, respectively\cite{pirjol}.
Using input values of $f_B = 216\mev$~\cite{ref:lattice},
$|V_{ub}| = 4.31\times10^{-3}$~\cite{pdg2006}, $\tau_{B} = 1.638\ps$~\cite{pdg2006}, and
$m_{b} = 4.20\gev$~\cite{pdg2006}, our
$90\%$ C.L. Bayesian limits on \delbf correspond to
values of $\lambda_B$ of 
$>669\mev$ and $>591\mev$, 
for the choice of the flat amplitude and flat BF priors, respectively.

Given a theoretical model, a measurement of \delbf may be converted into
an estimate of the total BF.  In the model of Ref.~\cite{korchemsky}, 
the result of the joint fit corresponds to a BF of 
$(-0.6^{+3.0}_{-3.4}(\mbox{stat})^{+1.3}_{-1.4}(\mbox{syst}))\times10^{-6}$, 
and $90\%$ C.L. Bayesian upper limits of 
$3.8\times10^{-6}$ and $5.0\times10^{-6}$  
for the flat amplitude and flat BF
priors, respectively.

We thank D.~Pirjol for help on the signal model, and C.~S.\ Kim and Y.\ Yang for
advice on the $\eta$ mode.  We are grateful for the excellent luminosity and machine conditions
provided by our \pep2\ colleagues, 
and for the substantial dedicated effort from
the computing organizations that support \babar.
The collaborating institutions wish to thank 
SLAC for its support and kind hospitality. 
This work is supported by
DOE
and NSF (USA),
NSERC (Canada),
IHEP (China),
CEA and
CNRS-IN2P3
(France),
BMBF and DFG
(Germany),
INFN (Italy),
FOM (The Netherlands),
NFR (Norway),
MIST (Russia),
MEC (Spain), and
PPARC (United Kingdom). 
Individuals have received support from the
Marie Curie EIF (European Union) and
the A.~P.~Sloan Foundation.


\begin{thebibliography}{99}
\bibitem{pirjol}
The calculation of the partial BF and its dependence on $\lambda_{B}$, 
based on Ref.~[1], was performed by D.~Pirjol.
\bibitem{ccmodes}
Charge-conjugate modes are included implicitly.
\end{thebibliography}
\end{document}